\documentclass[sigconf]{acmart}

\renewcommand\footnotetextcopyrightpermission[1]{} % removes footnote with conference information in first column

\usepackage{booktabs} % For formal tables
\usepackage[nolist]{acronym} %Abbreviation
\usepackage{url}      % llt: nicely formatted URLs
\usepackage{csquotes} % for using the enquote command
\AtBeginEnvironment{quote}{\itshape} %italics for quotes

\usepackage{comment}
\usepackage{soul}
\renewcommand{\emph}[1]{\ul{#1}}
\usepackage{todonotes}
\usepackage{colortbl}
\usepackage{ulem}
\usepackage{ifthen}
\usepackage{amssymb}
\usepackage{supertabular}
\usepackage{multirow}
\usepackage{balance}

\newboolean{showcomments}
\setboolean{showcomments}{true} % toggle to show or hide comments
\ifthenelse{\boolean{showcomments}}
{\newcommand{\nb}[2]{
		\fcolorbox{black}{yellow}{\bfseries\sffamily\scriptsize#1}
		{\sf\small$\blacktriangleright$\textit{#2}$\blacktriangleleft$}
	}
	
}
{\newcommand{\nb}[2]{}
	
}

\newenvironment{widequotation}{\list{}{\listparindent 0em \itemindent\listparindent
	\leftmargin 10pt	\rightmargin 0pt \parsep 0pt plus 1pt}\item\relax}
{\endlist}
\def\signed#1{{\leavevmode\unskip\nobreak\hfil\penalty50\hskip2em
		\hbox{}\nobreak\hfil\raise-1pt\hbox{#1}%
		\parfillskip=0pt \finalhyphendemerits=0 \endgraf}}
\newsavebox\mybox

\setcopyright{none}
\acmDOI{00.000/000_0} %TODO

\acmISBN{000-0000-00-000/00/00} %TODO

\acmConference[ARXIV'18]{Work in Progress Submission to Arxiv}{January 2018}{Gothenburg, Sweden} 
\acmYear{2018} %TODO
\copyrightyear{2018} %TODO
\begin{document}
\title{Ethical and Social Aspects of Self-Driving Cars}
%\titlenote{Produces the permission block, and copyright information}
%\subtitle{Extended Abstract}
%\subtitlenote{The full version of the author's guide is available as \texttt{acmart.pdf} document}

%TODO: CHECK ADDRESSES

\author{Tobias Holstein}
%\authornote{Test}
\orcid{0000-0001-6020-1785}
\affiliation{%
  \institution{M{\"a}lardalen University}
  \streetaddress{H{\"o}gskoleplan 1}
  \city{V{\"a}ster{\r{a}}s} 
  \country{Sweden} 
  \postcode{721 23}
}
\email{tobias.holstein@mdh.se}

\author{Gordana Dodig-Crnkovic, Patrizio Pelliccione}
%\authornote{Test.}
\orcid{0000-0001-9881-400X}
\affiliation{%
  \institution{Chalmers University of Technology | University of Gothenburg}
  \streetaddress{Maskingr{\"a}nd 2}
  \city{Gothenburg} 
  \country{Sweden}
  \postcode{412 96}
}
\email{[gordana.dodig-crnkovic,patrizio]@chalmers.se}

%\author{Patrizio Pelliccione}
%\authornote{Test}
%\orcid{0000-0002-5438-2281}
%\affiliation{%
%  \institution{Chalmers University of Technology | University of Gothenburg}
%  \streetaddress{Maskingr{\"a}nd 2}
%  \city{Gothenburg} 
%  \country{Sweden}
%  \postcode{412 96}
%}
%\email{patrizio.pelliccione@gu.se}

% The default list of authors is too long for headers.
%\renewcommand{\shortauthors}{T. Holstein et al.}

%Abbreviations
\begin{acronym}
    \acro{ECU}{Electronic Control Unit}
\end{acronym}

\begin{abstract}
As an envisaged future of transportation, self-driving cars are being discussed from various perspectives, including social, economical, engineering, computer science, design, and ethics. On the one hand, self-driving cars present new engineering problems that are being gradually successfully solved. On the other hand, social and ethical problems are typically being presented in the form of an idealized unsolvable decision-making problem, the so-called trolley problem, which is grossly misleading. We argue that an applied engineering ethical approach for the development of new technology is what is needed; the approach should be applied, meaning that it should focus on the analysis of complex real-world engineering problems. Software plays a crucial role for the control of self-driving cars; therefore, software engineering  solutions should seriously handle ethical and social considerations. 
In this paper we take a closer look at the regulative instruments, standards, design, and implementations of components, systems, and services and we present practical social and ethical challenges that have to be met, as well as novel expectations for software engineering.
%\footnote{This is an abstract footnote}
\end{abstract}

%
% The code below should be generated by the tool at
% http://dl.acm.org/ccs.cfm
% Please copy and paste the code instead of the example below. 
%
%TODO: Generate CCS for D.2 SOFTWARE ENGINEERING, D.2.9 Management, K.4.1 Public Policy Issues [ethics], K.4.2 Social Issues, I.2.9 Robotics [autonomous vehicles]
%
%
\begin{CCSXML}
<ccs2012>
 <concept>
  <concept_id>10010520.10010553.10010562</concept_id>
  <concept_desc>Computer systems organization~Embedded systems</concept_desc>
  <concept_significance>500</concept_significance>
 </concept>
 <concept>
  <concept_id>10010520.10010575.10010755</concept_id>
  <concept_desc>Computer systems organization~Redundancy</concept_desc>
  <concept_significance>300</concept_significance>
 </concept>
 <concept>
  <concept_id>10010520.10010553.10010554</concept_id>
  <concept_desc>Computer systems organization~Robotics</concept_desc>
  <concept_significance>100</concept_significance>
 </concept>
 <concept>
  <concept_id>10003033.10003083.10003095</concept_id>
  <concept_desc>Networks~Network reliability</concept_desc>
  <concept_significance>100</concept_significance>
 </concept>
</ccs2012>  
\end{CCSXML}

\ccsdesc[500]{Computer systems organization~Embedded systems}
\ccsdesc[300]{Computer systems organization~Redundancy}
\ccsdesc{Computer systems organization~Robotics}
\ccsdesc[100]{Networks~Network reliability}

%\keywords{ACM proceedings, \LaTeX, text tagging}
\keywords{Self-Driving Cars, Autonomous Cars, Trolley Problem, Decision Making, Ethics, Social Aspects, Software Engineering, Challenges}

\maketitle

\section{Introduction}

%\patrizio{command for my comments}
%\gordana{command for Gordana comments}
%\tobias{command for Tobias comments}
%\todo{For todo}
%\ins{Something to be added}\\
%\del{something to be deleted}\\
%\chg{something to be changed into}{that text}\\
%\ugh{something strange}\\

Increasingly, prototypical self-driving vehicles are participating in public traffic~\cite{Persson2014} and are planned to be sold starting in 2020~\cite{Toyota2015,Stoll_WSJ_2016}. Public awareness and media coverage contribute to a manifold of discussions about self-driving vehicles. This is currently amplified through recent accidents with autonomous vehicles~\cite{Tesla2016_tragicloss,Dolgov2016}. 

Software is playing a key role in modern vehicles and in self-driving vehicles. Gigabytes of software run inside the \acp{ECU}, which are small computers embedded in the vehicle. The number of \acp{ECU} has grown in the last 20 years from 20 to more than 100. Software in cars is growing by a factor of 10 every 5 to 7 years, and in some sense car manufacturers are becoming software companies~\cite{PELLICCIONE201783}. These novelties ask for a change on how the software is engineered and produced and for a disruptive renovation of the electrical and software architecture of the car, as testified by the effort of Volvo Cars \cite{PELLICCIONE201783}.

Moreover, self-driving vehicles will be connected with other vehicles, with the manufacturer cloud, e.g., for software upgrades, with Intelligent Transport Systems (ITS), Smart Cities, and Internet of Things (IoT). Self-driving vehicles will combine data from inside vehicle with external data coming from the environment (other vehicles, the road, signs, and the cloud). In such a scenario, different applications will be possible: smart traffic control, better platooning coordination, and enhanced safety in general.
However, the basic assumption is that future self-driving connected cars must be socially sustainable.
A typical discussion about ethical aspects of self-driving cars starts with ethical thought experiment, so called \enquote{trolley problem} described in~\cite{Foot1967} and~\cite{Wintersberger2017}, that has been discussed in number of articles in IEEE~\cite{7948873,Goodall2016,Ackerman2016}, ACM~ \cite{McBride:2016:EDC:2874239.2874265,Kirkpatrick:2015:MCD:2808213.2788477,Frison:2016:FPT:3004323.3004336}, Scientific American~\cite{Greenemeier2016,Deamer2016,Kuchinskas2013}, Science~\cite{Bonnefon2016,Greene2016_1514}, other high-profile journals \cite{2016arXiv160608813G_GoodmanFlaxman, Coca-Vila2017, goodall2014vehicle}, conference workshops~\cite{Riener:2016:WEI:3004323.3005687,Alavi:2017:DCW:3064857.3079155} and other sources \cite{MoralMachine2016,Mooney2016,Achenbach2015,Shashkevich2017}. Here is the general scenario being discussed:

\textit{
A self-driving vehicle drives on a street with a high speed. In front of the vehicle a group of people suddenly blocks the street. The vehicle is too fast to stop before it reaches the group. If the vehicle does not react immediately, the whole group will be killed. The car could however evade the group by entering the pedestrian way and consequently killing a previously not involved pedestrian. The following alternations of the problem exist: (A) Replacing the pedestrian with a concrete wall, which in consequence will kill the passenger of the self-driving car; (B) Varying the personas of people in the group, the single pedestrian or the passenger. The use of personas allows including an emotional perspective~\cite{BleskeRechek2010}, e.g., stating that the single pedestrian is a child, a relative, a very old or a very sick human, or a brutal dictator, who killed thousands of people.}

Even though the scenarios are similar, the responses of humans, when asked how they would decide, differ~\cite{Bonnefon2016}. The problem is that the question asked has limited number of possible answers, which are all ethically questionable and perceived as bad or wrong. Therefore, a typical approach to this problem is to analyze the scenarios by following ethical theories, such as utilitarianism, other forms of consequentialism or deontological ethics~\cite{mackinnon2012ethics}. For example, utilitarianism would aim to minimize casualties, even if it means to kill the passenger, by following the principle: the moral action is the one that maximizes utility (or in this case minimizes the damage). Depending on the ethics framework, different arguments can be used to justify the decision. 

Applying ethical doctrines to analyze a given dilemma and possible answers can presently only be done by humans. How would self-driving cars solve such dilemmas? There are a number of publications that suggest to implement moral principles into algorithms of self-driving cars~\cite{Goodall2016,DENNIS20161,Dennis2014}. We find that this does not solve the problem, but it reassures that the solution is calculated based on a given set of rules or other mechanisms, moving the problem to engineering, where it is implemented.

It is worth to notice that the engineering problem is substantially different from the hypothetical ethical dilemma. While an ethical dilemma is an idealized constructed state that has no good solution, an engineering problem is always by construction such that it can differentiate between better and worse solutions. A decision making process that has to be implemented in a self-driving car can be summarized as follows. It starts with an awareness of the environment: Detecting obstacles, such as a group of humans, animals or buildings, and also the current context/situation of the car using external systems (GPS, maps, street signs, etc.) or locally available information (speed, direction, etc.). Various sensors have to be used to collect all required information. Gaining detailed information about obstacles would be a necessary step before a decision can be made that maximizes utility and/or minimizes damage. A computer program calculates solutions and chooses the solution with the optimal outcome. The self-driving car executes the calculated action and the process repeats itself.

The process itself can be used to identify concrete ethical challenges within the decision making by considering the current state of the art of technology and its development. In a concrete car both the parts of this complex system and the way in which it is created have a critical impact on the decision-making. This includes for instance the quality of sensors, code, and testing. We also see ethical challenges in design decisions, such as whether a certain technology is used because of its lower price, even though the quality of information for the decision making would be substantially increased if more expensive technology (such as sensors) would be used.

Since building and engineering of self-driving vehicle involve various stakeholders, such as software/hardware engineers, sales people, management, etc., we can also pose the following questions: does the actual self-driving car have a moral on its own or is it the moral of its creators? And who is to blame for the decision making of a self-driving car? In ~\cite{DodigCrnkovic2012} the argument is put forward that the systemic view must be used in case of socio-technological systems. Thus the problems in the system can originate or be a combination coming from inadequate solutions in various steps from requirements specification to implementation, testing, deployment maintenance, safety regulation and other normative support etc. 

Besides the self-driving vehicle itself, it is also important to address  yet another complex system: self-driving vehicles participating in public traffic among cars with human drivers. Therefore, it is important to investigate how self-driving vehicles are actually built, how ethical challenges are addressed in their design, production, and use and how certain decisions are justified. Discussing this before self-driving vehicles are officially introduced into the market, allows taking part in the setting and definition of ethical ground rules. McBride states that \blockquote{Issues concerning safety, ethical decision making and the setting of boundaries cannot be addressed without transparency}~\cite{McBride:2016:EDC:2874239.2874265}. We think that transparency is only one factor, as it is necessary to start further investigations and discussions. 

In order to give a more detailed perspective on the complex decision making process, we propose to create a conceptual ethical model that connects the different components, systems and stakeholders. It shows inter-dependencies and allows pinpointing ethical challenges that will be presented in the concluding recommendations.

Focusing on important ethical challenges that should currently be addressed and solved is an important step before ethical aspects of self-driving cars can actually be meaningfully discussed from the point of view of societal and individual stakeholders as well as designers and producers. It is important to focus not on abstract thought experiments but on concrete conditions that influence the behavior of self-driving cars and their safety as well as our expectations from them.

The paper is structured as follows. A short introduction to self-driving cars and their current state of the art is provided in Section~\ref{sec:SelfDrivingCarsBasics}, with the emphasis on the description of the decision making principles given in Section~\ref{sec:SelfDrivingCarsBasics:DecisionMakingProcess} and the role of software in Section~\ref{sec:SelfDrivingCarsBasics:ComplexityOfDecisionMaking}. Ethical and social challenges are addressed in Section~\ref{sec:EAofTC} regarding technical aspects, and Section~\ref{sec:EAofNONTC} addressing social aspects. Section~\ref{sec:LegislationStandardGuidelines} describes the current state of norms and standards, while conclusions and final remarks are  presented together with recommendations in Section~\ref{sec:Conclusions}.

\section{Self-Driving Cars Basics}
\label{sec:SelfDrivingCarsBasics}

The term "autonomous" could be ambiguous to some readers. It can be used to describe certain autonomous features or functions, such as advanced driver assistance systems, that for example assist the driver in keeping the lane or adjust to the speed of vehicles ahead. Those systems are designed to assist, but the driver is always responsible and has to intervene if critical situations occur. 

We use the term "self-driving" cars to avoid wrong interpretations of the terms "fully autonomous" or "driverless". Self-driving cars refer to cars that may operate self-driving without human help or even without a presence of human being. This means that the unoccupied car can drive from place A to B to pick up someone. This is the highest level of autonomy for cars and corresponds to the last level of five as defined by the Society of Automotive Engineers \cite{SAE2016} and United States National Highway Traffic Safety Administration (NHTSA), who, since September 2016, adopted SAE's classification with level 1 (no automation), level 2 (drive assistance), level 3 (partial automation), level 4 (conditional automation), and level 5 (full automation) \cite[p.9]{NationalHighwayTrafficSafetyAdministrationNHTSA2016}.

A concrete example is the self-driving Waymo car \cite{Waymo2017}, former known as the Google car \cite{Google2016}, a fully autonomous and self-driving vehicle.

\subsection{Decision Making Process in Self-Driving Cars}
\label{sec:SelfDrivingCarsBasics:DecisionMakingProcess}

\begin{figure}
%TODO: \includegraphics[height=1in, width=1in]{fly}
\centering
\includegraphics[width=1\linewidth]{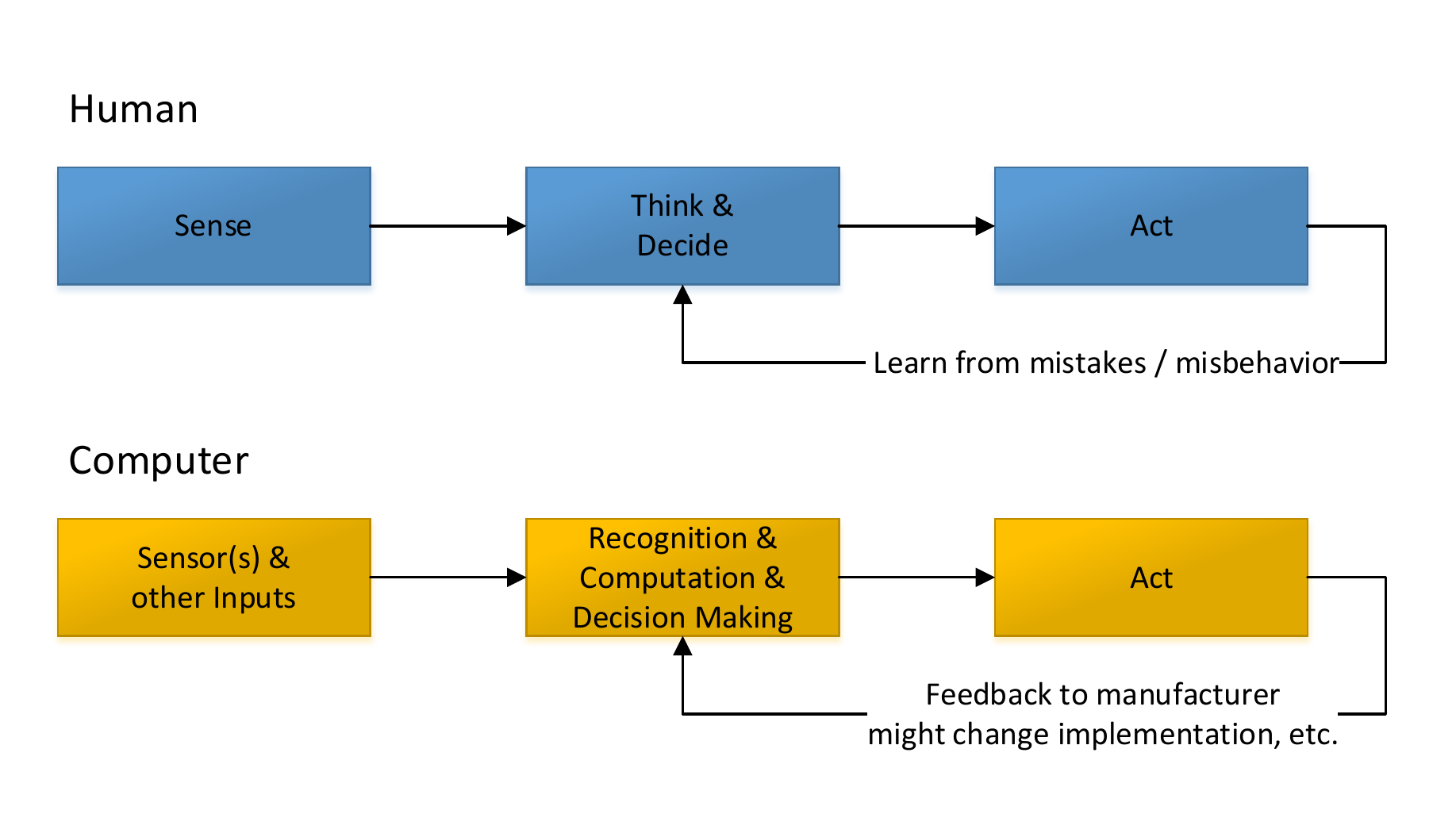}
\caption{Comparison of human and computers sense, think and act process (cf. \cite{Ghisio2016}) which we extended by adding a feedback loop}
\label{fig:ComparisonHumanComputerProcess}
\end{figure}

Developing self-driving cars that act without a driver means to replace a human, who today is performing the complex tasks of driving, with a computer system executing the same tasks. Figure~\ref{fig:ComparisonHumanComputerProcess} shows both variants and allows a comparison. 

%\begin{figure}
%	\centering
%	\includegraphics[width=1\linewidth]{artifacts.pdf}
%	\caption{Classification of artifacts}
%	\label{fig:artifacts}
%\end{figure}

There is an important difference in the feedback loop. While humans continuously learn, for example from their mistakes or misbehaviour, automotive software might be confined to slow updates. Approaches with self-adaptive software, such as machine learning approaches, which learns and reacts immediately, aim to overcome this constraint. Extraordinary road signs for example, which are new to the self-driving car's software, present a risk as they can pass unnoticed/uninterpreted, while they could be understood by a human through context/interpretation. Also unexpected and dangerous situations, like an attack or threat near or even against the vehicle might not be correctly interpreted by a self-driving car compared to a human.

\begin{figure}[h]
%TODO: \includegraphics[height=1in, width=1in]{fly}
\centering
\includegraphics[width=1\linewidth]{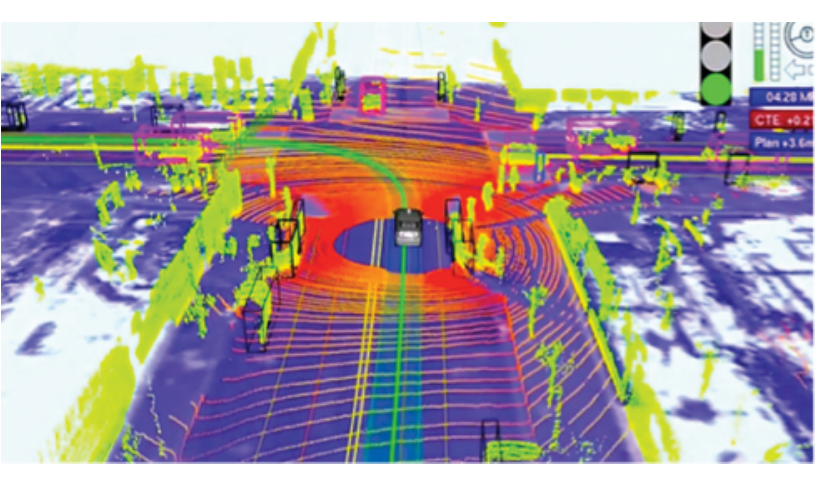}
\caption{Point cloud image of a vehicle approaching an intersection illustrates the complexity \cite{EarthImagingJournalEIJ:RemoteSensingSatelliteImages2012}}
\label{fig:PointCloudImageGoogle}
\end{figure}

Depending on the technology and the amount of sensors, the type and quality of information that is gathered differ. However, this extremely complex process might be difficult to imagine and in order to give an idea of what self-driving cars \enquote{see} we refer to the visualization depicted in Figure~\ref{fig:PointCloudImageGoogle}. It shows a rendered point cloud, based on the data gathered by a laser radar (LIDAR) mounted on the top of the vehicle. 

\subsection{Complexity of Decision Making and the Role of Software}
\label{sec:SelfDrivingCarsBasics:ComplexityOfDecisionMaking}

\begin{figure}
%TODO: \includegraphics[height=1in, width=1in]{fly}
\centering
\includegraphics[width=1\linewidth]{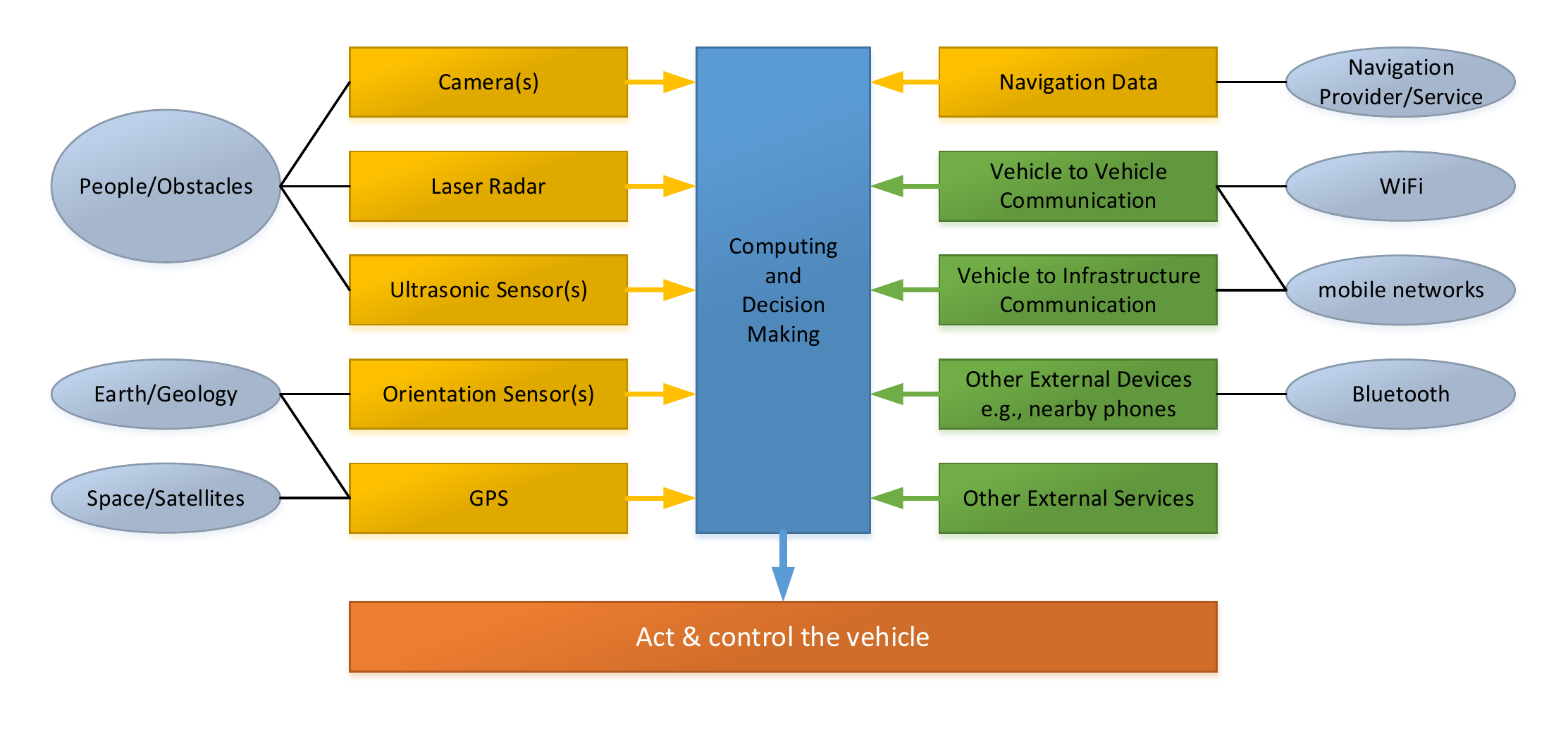}
\caption{Abstract representation of decision making in autonomous vehicles composed from various sources (cf.  \cite{EarthImagingJournalEIJ:RemoteSensingSatelliteImages2012,Waldrop2015,Waymo2017technology,Tesla2016_upgrade})}
\label{fig:DecisionMaking}
\end{figure}

The amount of sensors used to detect objects around the vehicle and its surrounding environment differs among car manufacturers. Figure~\ref{fig:DecisionMaking} shows an abstraction made to discuss the types of information used and how they relate to each other. 

Most of the functionality in the automotive domain is based on software~\cite{Broy2007}. Software is written by software engineers and at least for important components extensively tested to ensure their correct functioning. In self-driving cars software relies on different disciplines, such as computer vision, machine learning, and parallel computing, but also on various external services. It is a complex process to calculate a decision, and it is also difficult to test those against all possible real world scenarios~\cite{Waldrop2015}. 

One of the problems is that all calculations are based on an abstraction of the real world. This abstraction is an approximate representation of a real world situation and thus the decision making will create decisions for an imperfect world. This is a twofold problem, because the more information is available the better the decisions might be, but at the same time more interpretation and filtering might have to be used to get the data that actually is useful for the decision making. 

Engineers have to decide what kind of data to use, how reliable or trustworthy the data are and how to balance the different sources of information in their algorithms. Also different sensors have their specific limitations and to overcome those, a combination of multiple sensors might be used. The overall problem is usually referred to as sensor fusion. This problem is acerbated in the case of connected vehicles since data will come not only from the sensors of the car, but also from other vehicles, street infrastructure, etc. In this case other factors should be taken into account since it is not possible to have a perfect knowledge about the devices that are used to sense information and about their status.

Imagine heavy weather conditions, the navigation reports a street ahead, the radar is reporting a clear street, but the visual camera reports an obstacle straight ahead. How will this \enquote{equation} be solved and what will be the result? The wrong decision might lead to an accident, when important information of some sensors is disregarded and other sensors do not detect the obstacle or hazard in front of the vehicle~\cite{Tesla2016_tragicloss}. Car manufacturers are constantly improving and testing the recognition capabilities of their systems~\cite{Tesla2016_upgrade}. It is a multi-factor optimization task, which aims to find an optimal solution under consideration of costs, quality, and potential risk factors.

Some manufacturers are thinking to count miles covered without any accident, however this might be infeasible since a vehicle should cover around 11 billion of miles to demonstrate with 95\% of confidence and 80\% power that autonomous vehicle failure rate is lower than the human driver failure rate \cite{KALRA2016182}. Moreover, this calculation holds if the software within the car does not change over time. Nowadays, manufacturer are increasingly interested in continuous integration and deployment techniques that promise to update the software even after the vehicle has been sold and is on the street, like a common smart-phone. However, changing even a single line of code might require to starting counting from 0 the number of covered miles.

\section{ETHICAL ASPECTS OF THE TECHNICAL CHALLENGES IN SELF-DRIVING CARS}
\label{sec:EAofTC}

In the following, we will discuss ethical deliberations surrounding the autonomous vehicle, including involved stakeholders, technologies, social environments, and costs vs. quality. The multifaceted and complex nature of reality emphasizes again the importance to look broader instead of focusing on single ethical dilemma like the trolley problem.

\subsection{Safety}
\label{sec:EAofTC:Safety}

Safety is the most fundamental requirement of autonomous cars. The central question is then: how should a self-driving car be tested? What guidelines should be fulfilled to ensure that it is safe to use? There are several standards, such as the ISO 26262, that specify the safety standard for road vehicles. For self-driving cars standards are under development, based on experiences being made. 
Google Car tests show one million kilometres without any accident, is this a measurement to certify its software? As we discussed above, this should not be a reasonable assurance for safety. Should a self-driving car make a driver licence as suggested in \cite{McBride:2016:EDC:2874239.2874265}? How would that work? 

The source code of autonomous cars are typically commercial and not publicly available. One possibility to assure code correctness via independent control. Should there be an independent organization to check those? But could it actually be checked? Who else than the developers at a car manufacturer or supplier will understand such a complex system? 

An alternative route seems to be preferred by legislators - instead of control of the software which is in the domain of the producers, legislation focus on behaviour that is being tested, based on the "Proven in Use" Argument.

Testing of present-day cars should demonstrate the compliance of their behaviour with legislative norms \cite{DepartmentofMotorVehiclesStateofCalifornia}. Disengagements, accidents and reaction times based on data released in 2016 from the California trials are discussed in \cite{10.1371/journal.pone.0168054}.

In the case of the software of the car will evolve even when the vehicle is already on the street, testing should account for this new challenge.

When it comes to hardware and hardware-software systems, there have been discussions about the prices of laser radars compared to cameras or ultra-sonic sensors. Laser radars are very expensive, but deliver high quality data in diverse weather conditions. Ultra-sonic sensors or cameras are less accurate and sensitive under weather conditions like rain. 
Should a car manufacturer choose a cheap over an expensive sensor, even if this raises the likelihood of errors/faults/accidents? In advanced driving assistance systems, the driver would take over, if a critical situation could not be handled by the system. What happens in self-driving cars? Will the car just stop and wait until the rain is over? Will passengers be able and allowed to intervene? Under which conditions? Would it be required to have a driving licence for a self-driving car? Or would the police have a possibility to intervene, and in what way, when a car behaves inadequately or even dangerously? Also would the police even have the possibility to stop a self-driving car that is behaving correctly, with the sole purpose of checking the passengers?

The economic aspects might be seen as the highest priority. Using cheap equipment might lead to wrong decision-making and in a self-driving car, it would be impossible to interfere with the decisions made. Assuming that wrong decision may lead to a loss of human lives or property, having chosen a cheap component could therefore be ethically unacceptable.

Learning from experience is the most important basis for improvement of safety in self-driving cars. This is for instance envisioned by the CEO of Tesla, Elon Musk, in the Tesla's second 10 year master plan \enquote{part deux}, where the third element of the four major elements is: develop a vehicle self-driving capability that is 10x safer than manual via massive fleet learning\footnote{\url{https://www.tesla.com/blog/master-plan-part-deux}}.

\subsection{Security}
\label{sec:EAofTC:Security}

For autonomous cars, security is of paramount importance, and software security is a fundamental requirement. As an indication of the development we mention that in August 2017 UK's Department for Transport, published the document \enquote{Key principles of vehicle cyber security for connected and automated vehicles} \cite{DepartmentforTransportDfT2017}. It is built on the following eight basic principles:

\begin{enumerate}
\item Organizational security is owned, governed, and promoted at board level;
\item Security risks are assessed and managed appropriately and proportionately, including those specific to the supply chain;
\item Organizations need product aftercare and incident response to ensure systems are secure over their lifetime;
\item All organizations, including sub-contractors, suppliers, and potential 3rd parties, work together to enhance the security of the system;
\item Systems are designed using a defence-in-depth approach;
\item The security of the software is managed throughout its lifetime;
\item The storage and transmission of data is secure and can be controlled;
\item The system is designed to be resilient to attacks and respond appropriately when its defences or sensors fail.
\end{enumerate}

Similar documents are mentioned, such as Microsoft Security Development Lifecycle (SDL), SAFE Code best practices, OWASP Comprehensive, lightweight application security process (CLASP), and HMG Security policy framework~\cite{DepartmentforTransportDfT2017}.

There have been a number of attacks at car systems and sensors (e.g., LIDAR and GPS) that were used to manipulate the cars behaviour. Attacks might be inevitable, but should there be a minimum security threshold to allow a self-driving car to be used? This leads to another question: How secure must the systems and the connections be? 

In aircrafts \enquote{black boxes} are used to determine what happened after a crash. Should this be also a part of a self-driving car? 

What about security issues and software updates? Should a self-driving car be allowed to drive, when it does not have the latest software version running? What about bugs in the new software?

Should the vehicle be connected or should the vehicle be completely disconnected? On one side, the most secure system is the one that is disconnected from the network. On the side, it would be unethical to do not deploy immediately new software or a new version of the software on the car if there is evidence that the new update will fix important problems that might endanger human lives. 
In order to enable the massive fleet learning and to do the software update, connectivity is needed. Moreover, connected vehicles might receive information from other systems that will enhance the understanding of the reality thus opening new and promising safety scenarios. Imagine, for instance, a pedestrian on a side of a building, totally invisible to the instrumentations of the car, that is approaching a cross and that will most probably have an impact with the vehicle\footnote{\url{https://www.youtube.com/watch?v=w0rPQpjZhxg}}.

\subsection{Privacy}
\label{sec:EAofTC:Privacy}

The more information taken into consideration for the decision making, the more it might interfere with data and privacy protection. For example, a sensor that detects obstacles, such as human beings in front of the car is based on visual information. Even the use of a single sensor could invade privacy, if the data is recorded/reported and/or distributed without the consent of the involved people. The general question is: How much data is the car supposed to collect for the decision making? Who will access those data? When will these data be destroyed?

What about using active signals by devices people carry around to detect moving obstacles in front or near the car? What about people who do not carry such devices? Would they more likely be hit by the self-driving car, because they were not \enquote{present enough} in the data?

And how much data is actually used for evaluation? Is it anonymous? Does it contain more data than \enquote{just} the position of a human? Can it be connected to other types of data like the phone number, the bank account, the credit cards, personal details, or health data?

Those and similar questions are met by legislation such as Regulation (EU) 2016/679 of the European Parliament and of the Council (the General Data Protection Regulation) setting a legal framework to protect personal data \cite{EuropeanUnion2016}, and discussed in \cite{doi:10.1093/idpl/ipx005}.

\subsection{Trust}
\label{sec:EAofTC:Trust}

Trust is an issue that appears in various forms in autonomous cars e.g. in production (when assembled, trust is the requirement for both hardware and software components) as well as in use of the car. A human might define where the car has to go, but the self-driving car will make the decisions how to get there, following the given rules and laws. However the self-driving car might already distribute data like the target location to a number of external services, such as traffic information or navigation data, which are used in the calculation of the route. But how trustworthy are those data sources (e.g., GPS, map data, external devices, other vehicles)? 

In regard of the used sensors and hardware, the question is, how trustworthy are those? How can trust be implemented, when so many different systems are involved?

\subsection{Transparency}
\label{sec:EAofTC:Transparency}

The transparency is of central importance for many of the previously introduced challenges. Without transparency none of them could be analyzed, because the important information would be missing. \blockquote{Transparency is a prerequisite for ethical engagement in the development of autonomous cars. There can be nothing hidden, no cover-ups, no withholding of information} \cite{McBride:2016:EDC:2874239.2874265}. It is a multi-disciplinary challenge to ensure transparency, while respecting e.g., copyright, corporate secrets, security concerns and many other related topics. 

How much should be disclosed, and disclosed to whom? The car development ecosystem includes many other companies acting as suppliers that produce both software and hardware components. Should the entire ecosystem be transparent? Also to whom should it be transparent? How to manage the intellectual property rights? Some initial formulations are already present in the current policy documents and initial legislative that will be discussed later on.

Declaration of Amsterdam \cite{GovernmentNL2017} lists among the objectives \enquote{to adopt a \enquote{learning by experience} approach, including, where possible, cross-border cooperation, sharing and expanding knowledge on connected and automated driving and to develop practical guidelines to ensure interoperability of systems and services}.

Goodman and Flaxman in \cite{2016arXiv160608813G_GoodmanFlaxman} present EU regulations on algorithmic decision-making and a \enquote{right to explanation} that is the right for user to ask for an explanation of an algorithmic (machine) decision that was made about them.
The Department of Motor Vehicles provides the law requirements \cite{DepartmentofMotorVehiclesStateofCalifornia} \enquote{Under the testing regulations, manufacturers are required to provide DMV with a Report of Traffic Accident Involving an Autonomous Vehicle (form OL 316) within 10 business days of the incident}. The list of all incidents can be found in \cite{DepartmentofMotorVehiclesStateofCalifornia_OL316Reports}.

\subsection{Reliability}
\label{sec:EAofTC:Reliability}

One of the basic questions is: How reliable is the cell network? What if there is no mobile network available? What if sensor(s) fail? Should there be redundancy for everything? Is there a threshold that determines when the car is reliable, e.g., when two out of four sensors fail?  

In connected vehicles there are different levels that should be considered for reliability purposes. First the diagnostic of the vehicle that might be subject to failures. Then, the vehicle sensors that enable the vehicle to sense the surrounding environment of the vehicle. Finally, the data coming from external entities, like other vehicles and road infrastructures. Reliability approaches should consider all these levels.

\subsection{Responsibility and Accountability}
\label{sec:EAofTC:ResponsibilityAndAccountability}

In the case of autonomous cars responsibility will obviously be redefined. The question is how will responsibility be defined in case of incidents and accidents. Regarding ethical aspects of responsibility, a lot can be learned from the existing Roboethics and the debate about responsibility in autonomous robots, e.g., \cite{Dodig-Crnkovic:2008:SMR:1566864.1566888}. This is still an open problem even though important steps forward are being made by legislators, such as mentioned \enquote{Key principles of vehicle cyber security for connected and automated vehicles} \cite{DepartmentforTransportDfT2017}.

\subsection{Quality Assurance Process}
\label{sec:EAofTC:QualityAssuranceProcess}

Detailed Quality assurance programs covering all relevant steps must be developed in order to ensure high quality components. The question is also how is the decision making to be implemented. How to ensure overall quality of the product? What about the lifetime of components? How will maintenance be organized and quality assured?
When car manufacturers follow a non-transparent process of software engineering, how could anyone make sure that the car follows a certain ethical guideline? Whose responsibility will it be that car software follows ethical principles?

One part of the Quality Assurance (QA) process regards assembling of components. All parts of a vehicle are designed, fabricated and then assembled to the overall car. A standard non-autonomous premium vehicle today has more than 100 electronic control units that are responsible for the control of e.g., the engine, the wipers, the navigation system or the dashboard~\cite{PELLICCIONE201783}. We assume that for self-driving the number will be increased. Parts are usually built by not one, but a multitude of suppliers. This requires an extensive design and development process, which again involves various disciplines, such as requirements engineering, software engineering or project management. It is an overall extensive process which holds ethical questions and challenges. Thus, it is necessary to include ethical deliberations in the overall process but also in all sub processes. As it is stated in~\cite{Sapienza2016a}: \enquote{value-based ethical aspects, which today are implicit, should be made visible in the course of design and development of technical systems, and thus a subject of scrutiny}. 

Including ethics-aware decision making in all processes will help to make ethically justified decisions. This is important when it comes to questions: Which parts/components are used for a vehicle? Can we choose a cheaper component with less accuracy instead? Is the reliability of this part high enough for a self-driving car?

\section{Ethical Aspects of Social Challenges}
\label{sec:EAofNONTC}

%\subsection{Social Challenges}
\label{sec:EAofNONTC:SocialChallenges}

Self-driving cars will influence job markets, as for example for taxi drivers, chauffeurs or truck drivers. The perception of cars will change and cars might be seen as a service that is used for transportation. The idea of having a vehicle that is specialized for the specific use, e.g. off road, city road, long travels might become attractive. This might impact the business model of car manufacturer and their market.

This in itself poses ethical problems: what strategy should be applied for people loosing jobs because of the transition to self-driving cars? It is expected that the accident frequency will decrease rapidly, so car insurances may become less important. This may affect insurance companies in terms of jobs and the business. There is a historical parallel with process of industrialization and automatization, and there are experiences that may help anticipate and better plan for the process of transition.

\subsection{Stakeholders - General Public Interests}
\label{sec:EAofNONTC:Stakeholders}

Humans concerns must be taken into account in the decision making of self-driving cars.
Should there be an emergency button to allow the human to interfere with the decision making of the self-driving car? %} %\ins{
Putting the human back in the loop of decision making also inflicts with the autonomy of the system. Is it then truly self-driving? Giving passengers a choice to interfere with the decisions of the self-driving car puts the passenger back in charge, who would be responsible to press or \textit{not} press the button in all situations. %}. 
In the context of the self-driving car the computer decision might be better, but it might also be worse than human, because of possible errors~\cite{Eckstein2016}.
%\ins{However, a train operates in a partially controlled environment, then the train can stop without any consequence. This does not hold for cars. A car suddenly stopping might create accidents.}

Another perspective on the human interest is the granularity of the settings or configurations given to the user. How for example will a route be planned? %\ins{The user could mark the route on a virtual map or the route could always be determined by a computer system. The latter might choose a route that follows guidelines to ensure maximum efficiency and balanced traffic flows in an area. However, routes with maximum advertisement next to the street or with certain shops or attractions could also be preferred over others. }

In an extreme scenario self-driving cars might even avoid or reject to drive to a certain region or position. Would that be an interference with the freedom of choice, will passengers be informed about the reasons for such decisions? It is important to determine how much control the human should have, that will be taken into account when making design choices for a self-driving car.

\subsection{Possible New Selling Points}
\label{sec:EAofNONTC:SellingPoints}

The automotive industry has a highly competitive market. What will be the difference between buying a self-driving car of brand A compared to brand B? 

Taking away the primary and secondary tasks of driving, i.e. the driving controls, safety features, assistance etc., leaves only entertainment and comfort functions in control of the passengers, the former drivers. The interior becomes more important and factors that cannot be controlled less in the focus of the user. What will be the main buying criteria? Will it be the interior/exterior, speed (as often with traditional cars) or other new services? Will it be possible for the users of the car to choose the priorities in its decision-making? The latter is difficult, since decision choices~\cite{CarAndDriverTaylor2016} supporting the survival of passengers over other traffic participants by car manufacturers would have legal implications in most countries~\cite{Daimler2016pr}. The question is also who will own the cars. Will they become a service for individual users, and owned by companies? Buying criteria will be different depending on the ownership.

%\ins{Could the \enquote{character}, such as \enquote{aggressiveness} of a self-driving car be a criterion distinguishing between different cars? Will it be possible for a car to drive in the style of a racing driver? Will some cars be allowed to prioritize its passengers lives over all other participants in the traffic? It is worth to notice that this would be a false promise, as \enquote{trolley problem} presupposes that car indeed can predict and act perfectly, which we already argued is not the case in real life.}

Surveys based on hypothetical trolley problem scenarios show that people feel less attracted to buy a car that would sacrifice the passengers in order to save more human lives~\cite{Bonnefon2016}. Would that decision be left to car manufacturers? Existing policy documents do not seem to leave possibilities open for anti-social cars to be developed~ \cite{EthicsCommission2017pr,EthicsCommission2017b,Pillath2016,NHTSA2016PolicyUpdate,DBLP:journals/corr/CharisiDFLMSSWY17}.

Table 1 and 2 present the summary of ethical and social challenges with  recommendations (action points) grouped by requirement to be taken into account in policy-making as well as software design and development for self-driving cars.

\begin{table*}[t]
	%\vspace{-1.5em}
	\centering
	\caption{Summary of the technical challenges and recommendations grouped by requirement}
	%	\vspace{-8pt}
	\label{tab:Findings}
	\begin{small}
		\begin{supertabular}{%>{}p{.02\textwidth}|
		>{\raggedright}p{.105\textwidth}>{\raggedright}p{.4\textwidth}>{\raggedright}p{.42\textwidth}}
			\toprule
			 %& 
			 {\bf Requirements} & {\bf Challenges} & {\bf Recommendations}  \tabularnewline
			\midrule
			%\multirow{40}{*}{{\rotatebox[origin=r]{90}{Technical challenges}}} & 
			%~\vspace{.5cm}%\\
			\vspace{.1cm}Safety & \vspace{.1cm}
					%\begin{itemize}
					%\item 
					\noindent $\bullet$~Trade-off between safety and other aspects like economic aspects\\
					%\item 
					\noindent $\bullet$~Boundaries of autonomy of self-driving cars and human (passenger) interactions\\
					%\item 
					\noindent $\bullet$~Police control and possibility of intervention with self-driving cars\\
					%\item 
					\noindent $\bullet$~Systemic solutions to guarantee safety in organizations (regulations, authorities, safety culture) %\\
					%\end{itemize}
 			& \vspace{.1cm}
					%\begin{itemize}
					%\item 
					\noindent $\bullet$~Assure means to guarantee that safety is not sacrificed because of other aspects (this is not so different from today. This can happen also for the braking system of non-autonomous cars)\\
					%\item 
					\noindent $\bullet$~It should be specified how a self-driving car will behave in cases that the car is not able to deal autonomously. In the future passengers will be able to drive a car \\
					%\item 
					\noindent $\bullet$~There is the need of clarifying the relationship of police and self-driving cars\\
					%\item 
					\noindent $\bullet$~New techniques and standards are necessary to guarantee safety in self-driving cars that will continuously update their software \vspace{.2cm}
					%\end{itemize} 			
 			 \tabularnewline \hline %\cmidrule{2-4}
			%&
			%\\
			\vspace{.1cm}Security & \vspace{.1cm}
					%\begin{itemize}
					\noindent $\bullet$~Identification and declaration of minimal necessary security requirements, that work as a threshold for deployment of self-driving cars\\ %Minimum security requirements?\\
					\noindent $\bullet$~Security in systems and connections\\ % How secure must the systems and connections be?\\
					\noindent $\bullet$~Deployment of software updates \\%How will software updates be deployed?\\
					\noindent $\bullet$~Storing and using received and generated data in a secure way %How can data generated and received be used and stored in a secure way? 
					%\end{itemize}
			& \vspace{.1cm}
					%\begin{itemize}
					\noindent $\bullet$~Provide technical solutions that will guarantee minimum security under all foreseeable circumstances\\
					\noindent $\bullet$~Anticipate and prevent worst case scenarios regarding security breaches\\
					\noindent $\bullet$~Continuous learning process must be in place to provide active security\\
					\noindent $\bullet$~Assure accessibility of the data even in the case of accidents so that it could be analyzed and lessons learned. \vspace{.2cm}%\\
					%\end{itemize}								
 		     \tabularnewline \hline %\cmidrule{2-4}
			%& 
			%\\
			\vspace{.1cm}Privacy & \vspace{.1cm}
				%\vspace{-8pt}
				%\begin{itemize}
				%	\item 
				\noindent $\bullet$~Trade-off between privacy and data collection/recording
				%	\item 
				\noindent $\bullet$~Use of technology that detects humans near/around the car, even if those humans do not carry any kind of electronics  
				%\end{itemize}
			& \vspace{.1cm}
				%\vspace{-8pt}
				%\begin{itemize}
				%	\item 
				\noindent $\bullet$~Following/applying legal frameworks to protect personal data, such as Regulation (EU) 2016/679 of the European Parliament \cite{EuropeanUnion2016} (discussed in \cite{doi:10.1093/idpl/ipx005})\\
				%	\item 
				\noindent $\bullet$~Justify the use of collected data through a transparent decision making process \vspace{.2cm} %\\
				%\end{itemize}	
 			 \tabularnewline \hline %\cmidrule{2-4}
			%& 
			%\\
			\vspace{.1cm}Trust & \vspace{.1cm}
				%\begin{itemize}
				%	\item 
				\noindent $\bullet$~How trust between both software and hardware components of complex systems can be implemented is not clear
				%\end{itemize}
			& \vspace{.1cm}
				%\begin{itemize}
				%	\item 
				\noindent $\bullet$~Further research on how to implement trust across multiple systems \\
				%	\item 
				\noindent $\bullet$~Provide trusted connections between components as well as external services \vspace{.2cm} %\\
				%\end{itemize}
 			 \tabularnewline \hline %\cmidrule{2-4}
			%& 
			%\\
			\vspace{.1cm}Transparency & \vspace{.1cm}
               % \begin{itemize}
				
				%	\item 
				\noindent $\bullet$~Information disclosure, what and to whom\\
				%	\item 
				\noindent $\bullet$~Transparency of the ecosystem\\
				%	\item 
				\noindent $\bullet$~Management of intellectual property rights
				%\end{itemize}
			& \vspace{.1cm}
				%\begin{itemize}
				%	\item 
				\noindent $\bullet$~Ensure transparency and provide insight into decision making
				%	\item 
				\noindent $\bullet$~Actively share knowledge, gained by "learning through experience", to ensure the interoperability of systems and services.\\
				%		\item 
				\noindent $\bullet$~Transparency is a prerequisite for the herein introduced challenges, since it is the key to potentially undisclosed background information \vspace{.2cm} %\\
				%\end{itemize}
 			 \tabularnewline \hline %\cmidrule{2-4}
			%& 
			%\\
			\vspace{.1cm}Reliability & \vspace{.1cm}
				%\begin{itemize}
				%	\item 
				\noindent $\bullet$~Reliability of required networks and solution for the case when the network is unavailable\\
				%	\item 
				\noindent $\bullet$~Reliability of sensors, and need for redundancy\\
				%	\item 
				\noindent $\bullet$~Way to determine when a car is not reliable anymore \vspace{.2cm}%\\
				%\end{itemize}	
			& \vspace{.1cm}
				%\begin{itemize}
				%	\item 
				\noindent $\bullet$~Define different levels for reliability (diagnostics, vehicle input sensors, external services)\\
				%	\item 
				\noindent $\bullet$~Determine reliability for components and the overall car
				%\end{itemize}
 			 \tabularnewline \hline %\cmidrule{2-4}
			%& 
			%\\
		\vspace{.1cm}	Responsibility and Accountability & \vspace{.1cm}
				%\begin{itemize}
				%	\item 
				\noindent $\bullet$~Responsibility and accountability in case of incidents and accidents\\
				%	\item 
				\noindent $\bullet$~Responsibility that car software follows ethical principles \vspace{.2cm}%\\
				%\end{itemize}
			& \vspace{.1cm}
				%\begin{itemize}
				%	\item 
				\noindent $\bullet$~Consider research and learn from robotics, i.e. Roboethics \cite{Dodig-Crnkovic:2008:SMR:1566864.1566888}\\
				%	\item 
				\noindent $\bullet$~Support development of solutions, e.g. by contributing to existing approaches  \cite{DepartmentforTransportDfT2017}
				%\end{itemize} \\
 			 \tabularnewline \hline %\cmidrule{2-4}
			%& 
			%\\
			\vspace{.1cm}Quality Assurance (QA) Process & \vspace{.1cm}
				%\begin{itemize}
				 %   \item 
				 \noindent $\bullet$~Quality of components\\
				 %   \item 
				 \noindent $\bullet$~Quality of decision making\\
				 %   \item 
				 \noindent $\bullet$~Lifetime and maintenance\\
				%	\item 
				\noindent $\bullet$~Trade-offs between non-transparent processes and external QA control of adherence of ethical principles/guidelines \vspace{.2cm}%\\
					%\item How do engineers ensure high quality components?
					%\item How is the decision making implemented?
					%\item How to ensure a certain quality?
					%\item What about the lifetime of components?
					%\item How will maintenance be organized and quality assured?
					%\item When car manufacturers follow a non-transparent process of software engineering, how could anyone make sure that the car follows a certain ethical guideline?
					%\item Whose responsibility will it be that car software follows ethical principles? %<- BETTER IN Responsibility and Accountability section?
				%\end{itemize}                   
			& \vspace{.1cm}
				%\begin{itemize}
				%	\item 
				\noindent $\bullet$~Ethical deliberations must be included in the process of design and development of self-driving cars\\
				%	\item 
				\noindent $\bullet$~Ethics-aware decision making must be part of the process to ensure ethically justified decisions
				%\end{itemize}
			%\tabularnewline \cmidrule{1-4}
 			 \tabularnewline 
 			 %\bottomrule
 			 %\hline
		\end{supertabular}
	\end{small}
	
\end{table*}

\begin{table*}[t]
	%\vspace{-1.5em}
	\centering
	\caption{Summary of social challenges and recommendations grouped by requirement}
	%	\vspace{-8pt}
	\label{tab:FindingsNonTechnical}
	\begin{small}
		\begin{supertabular}{%>{}p{.02\textwidth}|
		>{\raggedright}p{.105\textwidth}>{\raggedright}p{.41\textwidth}>{\raggedright}p{.41\textwidth}}
			\toprule
			 %& 
			 {\bf Requirements} & {\bf Challenges} & {\bf Recommendations}  \tabularnewline
			\midrule
			%\multirow{15}{*}{{\rotatebox[origin=r]{90}{Non Technical challenges}}} &
			%\\ 
			\vspace{.1cm} Social challenges of disruptive technology & \vspace{.1cm}
				%\begin{itemize}
				%	\item 
				\noindent $\bullet$~Handling job losses (e.g., taxi/truck drivers, traditional mechanics, insurance agents, etc.) \\
				%	\item 
				\noindent $\bullet$~Change of related markets and business models (e.g., car insurances, car manufacturers, etc.) \vspace{.2cm}%\\
				%\end{itemize}         
			& 	\vspace{.1cm}
				%\begin{itemize}
				%	\item 
				\noindent $\bullet$~Prepare strategic solutions for people loosing jobs \\
				%	\item 
				\noindent $\bullet$~Take advice/learn from historic parallels to industrialization and automatization
				%\end{itemize} 
			
			\tabularnewline \hline %\cmidrule{2-4}
			%& 
			%\\
			\vspace{.1cm}Stakeholders - general public interests & \vspace{.1cm}
				%\begin{itemize}
				%	\item 
				\noindent $\bullet$~Human concerns in the decision making of self-driving cars, e.g. possibility of interference with the decision making\\
				%	\item 
				\noindent $\bullet$~Freedom of choice hindered by the system (e.g. it may not allow to drive into a certain area) \\
				%\end{itemize} 
							%	\item 
				\noindent $\bullet$~Values and priorities: Ensure that general public values will be embodied in the technology, with interests of minorities taken into account\vspace{.2cm}% \\
				%\end{itemize}  
			& \vspace{.1cm}
				%\begin{itemize}
				%	\item 
				\noindent $\bullet$~Determine and communicate the amount of control a human has in context of the self-driving car\\
				%	\item 
				\noindent $\bullet$~The freedom of choice determined by regulations \\
				%\end{itemize}  
					%	\item 
				\noindent $\bullet$~Active involvement of stakeholders in the process of design and requirements specification
				%\end{itemize}      
 			 \tabularnewline \hline %\cmidrule{2-4}
  			%& 
  			%\\
  			\vspace{.1cm}Selling points & \vspace{.1cm}
				%\begin{itemize}
				%	\item 
				\noindent $\bullet$~Self-driving cars will have different buying criteria, depending on who will own the cars, big companies, social institutions such as municipalities or individual users, as they all have different preferences. Among those preferences environmental and sustainability criteria can be expected to play central role\\
				%	\item 
				\noindent $\bullet$~Existing policy documents don't seem to leave possibilities open for anti-social self-driving cars to be developed~ \cite{EthicsCommission2017b,Pillath2016,NHTSA2016PolicyUpdate,DBLP:journals/corr/CharisiDFLMSSWY17}. \vspace{.2cm}%\\
				%\end{itemize}         
			& \vspace{.1cm}
				%\begin{itemize}
				%	\item  
				%\end{itemize}   
				\noindent $\bullet$~Priorities and choices for the self-driving cars will result from the dialog between producers and future users \\
				\noindent $\bullet$~Ensure that existing and future policies and standards prevent the possibility of developing "anti-social" self-driving cars.
 			 \tabularnewline \hline %\cmidrule{2-4}
			%& 
			%\\
			\vspace{.1cm}Legislation, norms, policies and standards & \vspace{.1cm} 
				\noindent $\bullet$~Keeping legislation up-to-date with current level of automated driving, and emergence of self-driving cars \\
                \noindent $\bullet$~Creating and defining global legislation frameworks for the implementation of interoperable and development of increasingly automated vehicles\\
				\noindent $\bullet$~Defining the  guidelines that will be adopted by society for building self-driving cars. \\
				\noindent $\bullet$~Including ethical guidelines in design and development processes
			& \vspace{.1cm}
				\noindent $\bullet$~Car producers supporting and collaborating with legislators in their task to keep up-to-date with the current level of automated driving\\
				\noindent $\bullet$~Legislative support and contribution to global frameworks to ensure a smooth enrollment of the emerging technology\\
				\noindent $\bullet$~Include ethics in the overall process of design, development and implementation of self-driving cars. Ensure Ethics training for involved engineers \cite{Sapienza2016a, spiekermann2015ethical} \\
				\noindent $\bullet$~Establish and maintain a functioning socio-technological system in addition to functional safety standards.\vspace{.2cm}
 			 %\tabularnewline
 			  			 \tabularnewline %\hline
			%\bottomrule	
		\end{supertabular}
	\end{small}
	
\end{table*}

\section{Legislation, Standards, and Guidelines}
\label{sec:LegislationStandardGuidelines}

Present-day regulatory instruments for transportation systems are based on the assumption of human-driven vehicles. As the development and introduction of increasingly automated and connected cars proceed, from level 1 towards level 5 of automation, legislation needs constant updates \cite{EthicsCommission2017pr,EthicsCommission2017b,Pillath2016,NHTSA2016PolicyUpdate}. It has been recognized that present state regulatory instruments for human-controlled vehicles will not be adequate for self-driving cars: \enquote{existing NHTSA authority is likely insufficient to meet the needs of the time and reap the full safety benefits of automation technology. Through these processes, NHTSA will determine whether its authorities need to be updated to recognize the challenges autonomous vehicles pose}~\cite{NHTSA2016PolicyUpdate}. % NHTSA | National Highway Traffic Safety Administration

On 14 April 2016 EU member states endorsed the Declaration of Amsterdam \cite{GovernmentNL2017} that addresses legislation frameworks, use of data, liability, exchange of knowledge and cross-border testing for the emerging technology. It prepares a European framework for the implementation of interoperable connected and automated vehicles by 2019~\cite{EthicsCommission2017b}. It also considers roles of stakeholders:

 \begin{quote}
 Agreement by all stakeholders on the desired deployment of the new technologies will provide developers with the certainty they need for investments. For an effective communication between the technological and political spheres, categorization and terminology are being developed which define different levels of vehicle automation. \cite{Pillath2016} %\hfill
 \end{quote} %TODO CONFIRM REFERENCES, there seem to be number mixups
 
The question is thus how to ensure that self-driving cars will be built upon ethical guidelines, which will be adopted by society. The strategy is to rely on rigorously monitoring the behaviour of cars, while the details of implementation are within the responsibility of producers. That means among others that design and implementation of software should follow ethical guidelines. An example of ethical guidelines trying to think one step further is described in Sarah Spiekermann's book Ethical IT innovation \cite{spiekermann2015ethical}. 

The approach based on \enquote{learning by experience} and \enquote{Proven in use} argument \cite{GovernmentNL2017,NationalInstruments2014a,DBLP:journals/corr/SchabeB15} presupposes a functioning socio-technological assurance system that has strong coupling among legislation, guidelines, standards and use, and promptly adapts to lessons learned. Ethical analysis in~\cite{DodigCrnkovic2012, Thekkilakattil_7273594,Johnsen2017_7958474} addresses this problem of establishing and maintaining a functioning learning socio-technological system, while \cite{Johnsen2017_7958474} discusses why functional safety standards are not enough.

\section{Conclusions and final Remarks}
\label{sec:Conclusions}
Self-driving vehicles have been recognized as the future of transportation systems and will be successively introduced into the transport systems globally \cite{EthicsCommission2017pr,Pillath2016,NHTSA2016PolicyUpdate}. It is now the right time to start an investigation into the manifold of ethical challenges surrounding self-driving and connected vehicles \cite{EthicsCommission2017b}. As this new technology is being tested and gradually allowed on the roads under controlled conditions, the focus should be on the practical technological solutions and their social consequences, rather than on idealized unsolvable problems such as much discussed trolley problem. Conclusions reached from idealized problem discussions would be that it has no general solution under all circumstances. We can compare this situation with the development and introduction of first cars. If the developers of traditional driver-controlled cars asked about general responsibility of a human driver for traffic accidents before allowing them to enter traffic, they would never be accepted, as safety in general and under all circumstances cannot be guaranteed and indeed human factor is the major safety concern. This does not mean that we should not take care of the basic requirements like security, safety, privacy, trust etc. and social challenges in general including legislation and stakeholders interests. On the contrary, those real-world techno-social problems must be taken seriously.

Focusing on unsolvable idealized ethical dilemmas such as the trolley problem obfuscates true ethical challenges, starting with characteristics of the whole techno-social system supporting new technology, with the emphasis on maximizing learning, on machine-, individual-, and social-level \cite{DodigCrnkovic2012, DBLP:journals/corr/CharisiDFLMSSWY17}. The decision-making process and its implementation, which is central for the behaviour of a car, might internally use unreliable or insecure technology. Emerging technology of self-driving cars should follow ethical guidelines that stakeholders agree upon and should not be an autonomous black box with unknown performance. This poses new expectations, which affect software engineering that is involved in all its stages - from its regulatory infrastructure, to the requirements engineering, development, implementation, testing and verification \cite{Greene2016_1514,MoralMachine2016,Mooney2016,Ackerman2016,DBLP:journals/corr/CharisiDFLMSSWY17}. As software is integral part of a complex software-hardware-human-society system, we presented different types of issues that we anticipate will affect software engineering in the near future.

It is also the right time to discuss the border between what is technically possible in relation to what is ethically justifiable. Even if this might limit the possibilities, it will set the necessary ground for further developments. 
The discussion should cover different dimensions, namely business, technical, process, and organization. First of all, there is the need to open a serious trade off analysis between business needs and ethics. As discussed above we should certainly avoid to compromise safety because of business priorities, e.g. equipping the car with cheaper but unreliable sensors. 
For what concerns technical aspects, it is of key importance to include ethical thinking and reasoning into the design and development process of autonomous and self-driving vehicles. Ethical aspects should be considered in every phase of a software development process, from requirements, till testing, maintenance, and evolution. Architectural and design decisions should be taken through a process that includes ethics as first-class actor and by involving stakeholders that are relevant to this concern. These architectural and design decisions should then be embedded into the code that will run the self-driving vehicles and ensure its ethical aspects are taken care of.    
It is also necessary to enforce the transparency on those processes, so that independent evaluations become possible. Proper development processes, supported by suitable organization structure should promote and enable a serious discussion of ethics, and 
% Laws and regulations 
should emphasize the human interests, to make sure that the freedom of choice does not disappear in the new era of fully autonomous and self-driving vehicles.

\clearpage
\balance
\bibliographystyle{abbrv}
\bibliography{references} 

\end{document}